\documentclass[aps,pra,superscriptaddress]{revtex4}

\usepackage{latexsym}
\usepackage{amsmath}
\usepackage{graphicx}

\newcommand{\be}{\begin{equation}}
\newcommand{\ee}{\end{equation}}
\newcommand{\bd}{\begin{displaymath}}
\newcommand{\ed}{\end{displaymath}}

\begin{document}

\title{Self-Consistent Mean-Field Theory for Frustrated Josephson
Junction Arrays\footnote{in {\em Highlights in Condensed Matter
Physics}, A. Avella, R. Citro, C. Noce and M. Salerno eds., \\ AIP
Conference Proceedings {\bf 695}, pp. 164-175, AIP Press, New York
(2003), ISBN 0-7354-0167-5.}}

\author{F. P. Mancini}
\author{P. Sodano}
\affiliation{Dipartimento di Fisica and
 Sezione I.N.F.N.,  Universit\`a di
Perugia, \\ Via A. Pascoli, I-06123 Perugia, Italy}
\author{A. Trombettoni}
\affiliation{
I.N.F.M. and Dipartimento di Fisica, Universit\`a
di Parma, \\ parco Area delle Scienze 7A, I-43100, Parma, Italy}

\begin{abstract}
We review the self-consistent mean-field theory for
charge-frustrated Josephson junction arrays. Using
$\langle\cos{\varphi}\rangle$ ($\varphi$ is the phase of the
superconducting wavefunction) as order parameter and imposing the
self-consistency condition, we compute the phase boundary line between
the superconducting region ($\langle \cos{\varphi} \rangle \neq
0$) and the insulating one ($\langle \cos{\varphi} \rangle = 0$).
For a uniform offset charge $q=e$ the superconducting
phase increases with respect to the situation in which $q=0$.
Here, we generalize the self-consistent mean-field theory to
include the effects induced by a random distribution of offset charges
and/or of diagonal self-capacitances.
For most of the phase diagram, our results agree with the outcomes of
Quantum Monte Carlo simulations as well as with previous studies using
the path-integral approach.
\end{abstract}

\maketitle


\section{Introduction}

The first artificially fabricated Josephson junction arrays 
(JJA's) were realized twenty years ago at IBM \cite{voss82} 
as an effort to develop an electronics based on superconducting 
devices. Immediately after, it became clear that JJA's provided 
an ideal model to investigate classical phase transitions, 
frustration effects and relevant aspects of non-linear dynamics 
\cite{simanek94, fazio01}. JJA's are built by placing, 
on the sites of a lattice, islands of superconducting material 
coupled by Josephson junctions. The huge variety of behaviors 
of the system is rather simply described by the competition 
between the Josephson energy $E_J$ and the charging energy $E_C$: 
the former being responsible for the Josephson tunneling of Cooper 
pairs between the sites of the lattice, while the latter measures 
the effects of the electrostatic repulsion between Cooper pairs. The
superconductor-insulator transition typical of JJA's, for instance,
depends crucially on the ratio between these two energy scales.

In many situations, it is relevant to analyze the effect
of a background of external charges on the superconductor-insulator
transition of a quantum JJA. Offset charges
arise in experimental realizations of JJA's 
as a result of charged impurities or by the application 
of a gate voltage between the array and the ground.
In the former situation, offset charges are naturally
randomly distributed on the lattice while in the latter situation
they play the role of a sort of chemical potential and, then, their
distribution may also be uniform. Offset charges may be regarded
as effective charges $q_{i}$ located at the sites of the lattice: 
when $q_{i}\ne 2e$, they cannot be eliminated by Cooper
pair tunneling. In general, offset charges frustrate
the attempts of the system to minimize the energy of the charge
distribution of the ground state (for this reason they are also called
{\em frustration charges}). A large number of studies has by now been
devoted to the analysis of the effects induced by offset charges both
on the zero-temperature phase transition
\cite{roddick93,luciano9596,larkin97,choi01} and on the phase transition
at finite temperature \cite{bruder92,vanotterlo93,grignani00}.

In this paper we shall use the self-consistent mean-field theory (SCMFT)
to investigate the finite temperature phase diagram for the
self-charging (SC) model of JJA's \cite{simanek94}; in this model it is
assumed that the potential at site $i$ depends only on the charge at the
same site and, thus, the capacitance matrix describing the charge
effects of the array is diagonal. In particular, we shall investigate
situations in which offset charges (both uniform and random) are
present. Although quantum corrections may be relevant
for $d \le 2$, the SCMFT has the merit of providing a rather
intuitive and physically transparent approach to the analysis of some
general features of the superconductor-insulator transition in these
systems. Furthermore, the results we obtained are in very good
agreement with the outcome of recent numerical simulations
\cite{alsaidi03_2} and consistent with other analytical approaches
not relying on mean-field theory \cite{diamantini95}.

The plan of the paper is as follows: in Sec. II we review the SCMFT
for the SC model of quantum JJA's with a uniform distribution
of offset charges \cite{simanek94,grignani00}. 
We study the eigenvalue equation of the mean-field Hamiltonian and, 
for a uniform offset charge $q_{i}=e$, we show that there is
superconductivity for all values of $E_J/E_C$. In Sec. III we
point out how to extend the SCMFT to
situations in which capacitive disorder is present: one has to impose
the self-consistency condition with a double average, the quantum one
and the average over the disorder. The results are in agreement with the
ones obtained with the path-integral approach
\cite{mancini03,mancini03-1} and, at very low temperatures, are
consistent with the phase diagram obtained in Ref. \cite{fisher89}.
Section IV is devoted to our concluding remarks.

\section{Mean-Field Theory for JJA's with Offset Charges}

The Hamiltonian commonly used to describe Cooper pairs tunneling in
superconducting quantum networks defines the so-called quantum phase
model (QPM):
\begin{equation}
H=\frac{1}{2}\sum_{ij}(Q_i+q_i)C_{ij}^{-1}
(Q_j+q_j) -E_{J}\sum_{\left \langle ij\right\rangle }\cos
(\varphi_i-\varphi_j),
\label{QPM}
\end{equation}
where $\varphi_i$ is the phase of the superconducting order
parameter at the grain $i$. Its conjugate variable 
${n}_i \equiv Q_i /2e=-i
\partial/\partial \varphi_i$ (with $[\varphi_i, n_j]=i\:\delta_{ij}$)
describes the number of excess Cooper pairs on the $i$-th
superconducting grain and $C_{i j}$ is the capacitance matrix. The
symbol $\langle ij \rangle$ denotes a sum over nearest-neighbor
grains only.

The first term in the Hamiltonian (\ref{QPM}) determines the
electrostatic coupling between the Cooper pairs while the second term
describes the hopping of Cooper pairs between neighboring sites
($E_J$ is the Josephson energy). 
An external gate voltage $V_i$ provides a contribution to the
energy via the offset charge $q_i=\sum_j C_{i j} V_j$;
this external voltage can be either applied to the ground plane or
it may be induced by charges trapped in the substrate.
The former situation leads to the appearance of a uniform frustration
charge, while the latter naturally induces a random offset charge. In
this paper we shall limit our investigation only to the SC
model described by the Hamiltonian (\ref{QPM}) with
$C_{ij}=\delta_{ij}\: C_{ii}$. When all the $C_{ii}$'s are equal 
($ C_{ii}=C_0$), the charging energy $E_C$ is defined as 
$E_C=e^2 /2 C_0$.

With a uniform distribution of offset charges $q_{i}=q$,
the Hamiltonian of the array is given by:
\begin{equation}
H=4E_{C} \sum_{i}\left( i \frac{\partial}{\partial
\varphi_i}-\frac{q}{2e} \right)^2-  E_{J} \sum_{\langle i,j\rangle} \cos
(\varphi_i-\varphi_j).
\label{Ham}
\end{equation}
Mean-field theory for the SC model for quantum JJA's
was first used by Sim\`anek \cite{simanek79,simanek94}.
The approximation consists in replacing the Josephson coupling
on the link $i$-$j$ by an average coupling so that
$E_J\sum_{\langle i j\rangle}\cos(\varphi_i-\varphi_j) \approx
z E_J \langle\cos {\varphi}\rangle \sum_{i} \cos \varphi_i$, where
$z$ is the coordination number. Requiring the order parameter to be
real, leads to $\langle \sin{\varphi_{i}}\rangle=0$; it is also assumed
that $\langle \cos{\varphi}\rangle$ does not depend on the island index
$i$. In the mean-field approximation the Hamiltonian (\ref{Ham})is given
by a sum of single site Hamiltonians $H_{i}$ describing a quantum
particle in the potential $\cos{\varphi_i}$:
\begin{equation}
\label{MFA}
H_{MFA}=\sum_i H_i=\sum\limits_i \Bigl [-4 E_C
\frac{\partial^2}{\partial \varphi_i^2}-8i\frac{q}{2e} E_C
\frac{\partial}{\partial \varphi_i}+ 4 \left( \frac{q}{2e} \right)^2 E_C
- z E_J \langle\cos \varphi\rangle \cos \varphi_i \Bigr].
\end{equation}
The pertinent Schr\"odinger equation to be solved is then
\begin{equation}
\Bigl[-4 E_C \frac{d^2}{d\varphi^2}-8 i \frac{q}{2e}
E_C \frac{d}{d\varphi} +4 \left( \frac{q}{2e}\right)^2 E_C
-z J \langle\cos \varphi\rangle \cos \varphi \Bigr]
\psi_m(\varphi)=E_m \psi_m(\varphi).
\label{mf_se}
\end{equation}
Due to the periodicity of the phase $\varphi$, the eigenfunctions
should be $2\pi$-periodic functions of $\varphi$, i.e.,
\begin{equation}
\psi_n (\varphi)=
\psi_n (\varphi+2\pi).
\label{condizionefondamentale}
\end{equation}
Furthermore, since the Hamiltonian (\ref{QPM}) is invariant under the
shift $q \to q+2le$, where $l$ is an integer, it is relevant to
analyze only the situations corresponding to $q=0$ and $q=e$.

The order parameter $\langle\cos {\varphi}\rangle$ is evaluated
in terms of the eigenfunctions of Eq. (\ref{mf_se}) through the
self-consistency equation \begin{equation} \langle\cos
\varphi\rangle=\frac{\sum\limits_m e^{-E_m /k_B T} \langle
\psi_m\vert\cos \varphi \vert\psi_m\rangle}{\sum\limits_m e^{-E_m /k_B
T}}.
\label{3-64}
\end{equation}
From Eq. (\ref{3-64}), one immediately sees that, for high temperatures or
low $E_J$, only the solution $\langle\cos \varphi\rangle=0$ exists and,
thus, there is no superconductivity; for low temperatures or high $E_J$
instead, $\langle\cos \varphi\rangle\ne 0$ and the system as a whole
behaves as a superconductor. Solving the eigenvalue equation
(\ref{mf_se}) provides us with all the tools needed to investigate the
finite temperature phase diagram of the SC model of frustrated JJA's.
Defining $v=-z E_{J} \langle\cos \varphi\rangle/2 E_{C}$,
$\lambda_{m}^{\prime}=[E_{m}-4 E_{C} (q/2e)^{2}]/ E_{C}$, and $K=2 i
(q/2e)$, one finds 
\be
\frac{d^2 \psi_m}{d\varphi^2}+K \frac{d \psi_m}{d\varphi} +\Bigr(
\frac{\lambda_{m}^{\prime}}{4}-\frac{v}{2} \cos\varphi \Bigr) \psi_m=0 .
\label{h}
\ee
Equation (\ref{h}) is a Mathieu equation with a term proportional to a
first derivative: setting \begin{equation}
\psi_m (\varphi)= e^{-\frac{1}{2} K \varphi}
\rho_m (\varphi),
\label{i}
\end{equation}
one gets an equation for $\rho_m$, namely
\be
\frac{d^2 \rho_m}{d \varphi^2}+\Bigl(
\frac{\lambda_{m}^{\prime}}{4}-\frac{K^2}{4} -  \frac{v}{2} \cos \varphi
\Bigr) \rho_m=0.
\label{l}
\ee
If one sets $\lambda_m=E_m/E_C$ and $\varphi=2x$,
the eigenvalue equation (\ref{h}) is usefully recast in the standard
form of the Mathieu equation \cite{abramowitz64}:
\be
\frac{d^2 \rho_m}{dx^2}+(\lambda_m-2 v\cos 2x) \rho_m=0.
\label{eqmath1}
\ee
It is well known \cite{abramowitz64} that the Mathieu equation
admits the following periodic solutions:
\begin{enumerate}
\item $ce_{2n} (x,v)$, even solutions with period $\pi$
corresponding to the eigenvalues $a_{2n} (v)$;
\item $se_{2n+2} (x,v)$, odd solutions with period $\pi$
corresponding to the eigenvalues $b_{2n+2} (v)$;
\item $ce_{2n+1} (x,v)$, even solutions with period $2 \pi$
corresponding to the eigenvalues $a_{2n+1} (v)$;
\item $se_{2n+1} (x,v)$, odd solutions with period $2 \pi$
corresponding to the eigenvalues $b_{2n+1} (v)$.
\end{enumerate}
Since the eigenfunctions of the Schr\"{o}dinger equation (\ref{mf_se})
should satisfy the periodic boundary condition
(\ref{condizionefondamentale}), from Eq. (\ref{i}) one immediately sees
that one should treat differently the situations where $q/2e$ is integer
or half-integer. In fact, for integer $q/2e$, one has to consider only
$\pi$-periodic solution of Eq. (\ref{eqmath1}): in this way
$\rho_m(\varphi)$ are $2\pi$-periodic, which, in turn, leads to
$2\pi$-periodic $\psi_m(\varphi)=e^{-i l \varphi}
\rho_m(\varphi)$ (since $l$ is an integer). The solutions of Eq.
(\ref{eqmath1}) with period $\pi$ are the Mathieu eigenfunctions
$ce_{2n+2} (x)$, $se_{2n} (x)$ (with $n=0,1,\ldots$)
\cite{abramowitz64}. If, instead, $q/2e$ is half-integer, Eqs.
(\ref{condizionefondamentale}) and (\ref{i}) require the use of
the Mathieu eigenfunctions 
$ce_{2n+1} (x)$, $se_{2n+1} (x)$ (with $n=0,1,\ldots$) 
which are $\pi$-anti-periodic. Then
$\rho_m(\varphi)$ are $2 \pi$-anti-periodic and the eigenfunctions
$\psi(\varphi)=e^{-i l \varphi/2} \rho_m(\varphi)$ are $2\pi$-periodic.

Since the phase transition is expected to be second order
\cite{simkin96}, near the transition temperature, the order parameter
$\langle \cos\varphi \rangle$ and the parameter $v$ are small: this
allows one to use the expansion for small $v$ of the Mathieu functions 
\cite{abramowitz64}. As a result, one finds that, at the first order 
in $v$ and  apart from the phase factor $e^{-i\varphi/2}$, 
the normalized eigenfunctions of the Schr\"{o}dinger equation
(\ref{mf_se}) satisfying, for $q=e$, the condition
(\ref{condizionefondamentale}) are given by
\be
\begin{split}
\psi_1^e
(\varphi)&=\frac{1}{\sqrt{\pi}} \Bigl( \cos \frac{\varphi}{2}
-\frac{v}{8} \cos \frac{3 \varphi}{2} \Bigr)\\
\psi_1^o
(\varphi)&=\frac{1}{\sqrt{\pi}} \Bigl( \sin \frac{\varphi}{2}
-\frac{v}{8} \sin \frac{3 \varphi}{2} \Bigr),
\end{split}
\label{eigenfunctions1}
\end{equation}
and, for $n=1,2,\ldots$, by
\begin{equation}
\begin{split}
\psi_{2n+1}^e
(\varphi)&=\frac{1}{\sqrt{\pi}} \Bigl \{ \cos
\frac{(2n+1)\varphi}{2}-v  \Bigl[ \frac{\cos
\frac{(2n+3)\varphi}{2}}{4 (2n+2)}  -\frac{\cos
\frac{(2n-1)\varphi}{2}}{8n} \Bigr] \Bigr \}\\
\psi_{2n+1}^o (\varphi)&=\frac{1}{\sqrt{\pi}} \Bigl \{ \sin
\frac{(2n+1)\varphi}{2}-v\Bigl[ \frac{\sin
\frac{(2n+3)\varphi}{2}}{4 (2n+2)}  -\frac{\sin
\frac{(2n-1)\varphi}{2}}{8n} \Bigr] \Bigr \},
\end{split}
\label{eigenfunctions2}
\end{equation}
where the superscript $e$ ($o$) stands for $even$ ($odd$). The
corresponding energy eigenvalues are
\begin{equation}
\begin{split}
E_1^e &=E_{C}(1+v)/2,\\
E_1^o &=E_C (1-v)/2,\\
E_{2n+1}^e &=E_{2n+1}^o=E_C(2n+1)^2/2\\
n &=1,2,\ldots \quad \quad \quad \quad.
\end{split}
\label{eigenvalues}
\end{equation}
The expectation values
$\langle\psi_m\vert\cos \varphi \vert\psi_m\rangle$, 
at the order $v$, are given by
\begin{equation}
\begin{split}
\langle\psi_1^e\vert\cos \varphi\vert\psi_1^e\rangle &=\frac{1}{2}
-\frac{v}{8} \\
\langle\psi_1^o\vert\cos \varphi\vert\psi_1^o\rangle &=
-\frac{1}{2}-\frac{v}{8} \\
\langle\psi_{2n+1}^e\vert\cos
\varphi\vert\psi_{2n+1}^e\rangle &=
\langle\psi_{2n+1}^o\vert\cos\varphi\vert\psi_{2n+1}^o\rangle =
\frac{v}{8n(n+1)}\\
n &=1,2,\ldots \quad.
\end{split}
\label{expectation}
\end{equation}
Upon inserting the above eigenfunctions and eigenvalues in
Eq. (\ref{3-64}) and keeping only the terms  proportional to $v \sim
\langle\cos \varphi\rangle$, 
one obtains the following equation for the critical temperature $T_c$:
\begin{equation}
\frac{1}{\alpha}=g(q=e,y),
\label{r}
\end{equation}
where $y=k_B T_c/E_C$, $\alpha=z J/(4 E_C)$ and
\begin{equation}
g(q=e,y)=\frac{ \frac{4+y}{4y}
e^{-1/y}-\sum\limits_{n=1}^{\infty}
\frac{1}{1-4(n+1/2)^2} e^{-(4/y)(n+1/2)^2}}
{e^{-1/y}+\sum\limits_{n=1}^{\infty} e^{-(4/y)(n+1/2)^2}}. 
\end{equation}
From Eq. (\ref{r}) one easily shows that,
in the presence of charge frustration $\pm (2n+1) \> e$ 
on the lattice sites, for each value of $\alpha$, 
there is a insulator-superconductor transition. 
Indeed, $g(e,y) \to \infty$ for $y \to 0$ and $g(e,y) \to 0$ for
$y \to \infty$; also $dg/dy <0$ for all $y>0$.
It follows that Eq. (\ref{r}) has a unique solution for each value
of $\alpha$. Moreover, since $g\approx 1/y=1/\alpha$,
the critical temperature at which the transition occurs 
is given by $T_c \approx z E_{J}/4$.

\begin{figure}
\includegraphics[height=.2\textheight]{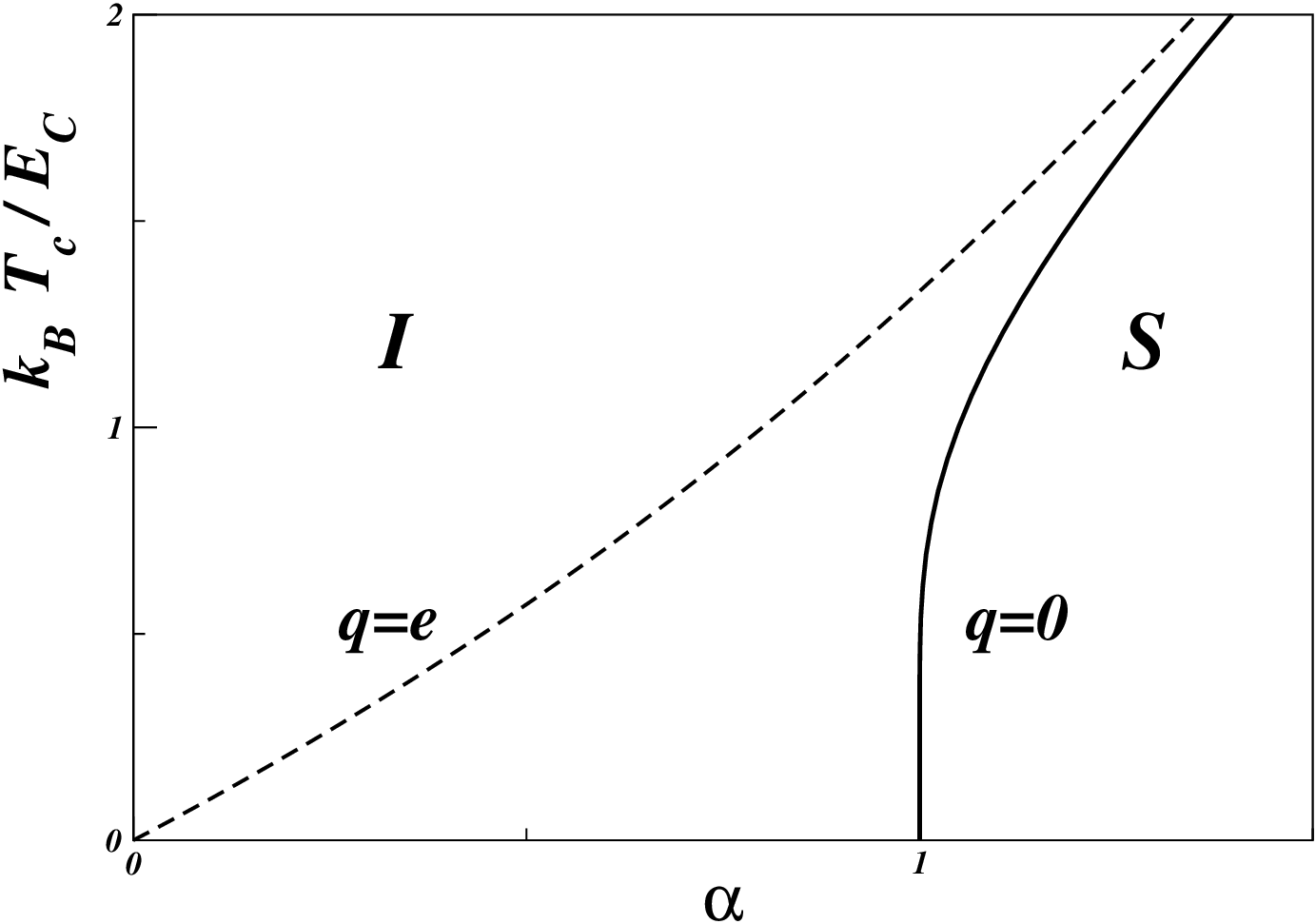}
\caption{Phase diagram for the SC model  without charge frustration
(solid line) and with charge frustration $q=e$ (dashed
line). $\alpha$ stands for the ratio $z E_J /4E_C$. The {\bf I} and
{\bf S} indicate, respectively, insulating and superconducting phase.}
\label{fig1}
\end{figure}

For $q=0$, the solutions of the Mathieu equation (\ref{eqmath1}) with
period $\pi$ are the Mathieu eigenfunctions $ce_{2n+2} (x)$, $se_{2n}
(x)$ (with $n=0,1,\ldots$): one finds
\begin{equation}
\frac{1}{\alpha}=g(q=0,y)
\label{r0}
\end{equation}
where
\begin{equation}
g(q=0,y)=\frac{1-2 \sum\limits_{n=1}^{\infty} \frac{e^{-4n^2/y}}{4 n^2-1}}
{1+2\sum\limits_{n=1}^{\infty} e^{-4n^2/y}}.
\end{equation}
From Eq. (\ref{r0}) one sees that for $\alpha >1$
the system has an insulator-superconductor transition, while for
$\alpha<1$  there is no evidence for a transition. In fact, $g(e,y) \to
\infty$ for $y \to 0$ and $g(e,y) \to 1$ for $y \to \infty$; also $dg/dy
<0$ for all $y>0$. Therefore, the self-consistency equation (\ref{3-64})
does not have solutions for $\alpha<1$ and it has a unique solution for
$\alpha>1$. For small values of $y$, Eq. (\ref{r0}) yields $1/\alpha
\approx (1 - 2 e^{-4/y} / 3 ) / (1 + 2 e^{-4/y})$, from which
\begin{equation}
\frac{k_B T_c}{E_C} \approx \frac{4}{\log \frac{2
(\alpha+3)}{3 (\alpha-1)}}. \label{3-77a}
\end{equation}
In Fig. \ref{fig1} we plot $T_c$ as a function of $\alpha$
for $q=0$ and $q=e$ using Eqs. (\ref{r}) and (\ref{r0}).
For $q=0$, one sees that there is no superconductivity for $\alpha<1$.
For $q=e$, superconductivity is attained for all the
values of $\alpha$: a uniform offset charge $q=e$ always {\em favors}
superconductivity.

It is worth noting that the eigenfunctions $\psi_m^{(0)}$ of the
Schr\"odinger equation (\ref{mf_se}) without the periodic potential are
\begin{equation}
\psi_m^{(0)}=\frac{1}{\sqrt{2 \pi}} e^{\pm i m \varphi/2}.
\label{onde_piane}
\end{equation}
These wavefunctions are also eigenfunctions of the number operator with
eigenvalues ${\cal N}=\langle\psi_m^{(0)}\vert -i \partial/\partial
\varphi \vert\psi_m^{(0)}\rangle=\pm \, m / 2$. When $q=0$, the wave
functions $\psi_{2n}^e$ and $\psi_{2n}^o$ are, respectively, the even
and odd combinations resulting from the splitting of the eigenfunctions
$\psi_{2n}^{(0)}$ and are related to the expectation value of the
half-integer number of Cooper pairs (${\cal N}=0,\pm 1,\pm 2,\ldots$).
On the other hand, when $q=e$, the expectation value of the number
operator on the eigenfunctions $\psi_{2n+1}^e$ and $\psi_{2n+1}^o$ is
half-integer and is equal to $1/2$ in the ground state. Therefore, an
offset charge $q=e$ favors the Cooper pairs tunneling, making possible
the insulator-superconductor transition also when $E_{C} \gg E_{J}$.

If one should use both periodic and anti-periodic solutions, 
the general solution of Eq. (\ref{mf_se}) would not have a
definite periodicity and, consequently, the charges $n_i$ 
would take any value; this situation is expected to be relevant 
in the description of continuous flows of current due, for instance, 
to ohmic shunt resistances \cite{likharev85,schon88}. 
Unless there is dissipation, the use of both periodic and 
anti-periodic solution is unwarranted; however if, in the
self-consistency equation (\ref{3-64}), one should include also the
$2\pi$-anti-periodic eigenfunctions \cite{simanek79}, one would find - for small
critical temperatures - the following equation
\begin{equation}
1=\alpha \frac{1+(\frac{2}{y}+\frac{1}{2})e^{-1/y}-\frac{2}{3}
e^{-4/y}}{1+2e^{-1/y}+2e^{-4/y}}. 
\label{equazionerientrante}
\end{equation}
Equation (\ref{equazionerientrante}) for $\alpha$ less than a critical value
$\alpha_c$ ($\alpha_c \approx 0.79$) does not have solutions, for
$\alpha_c<\alpha<1$ has two solutions and for $\alpha >1$ has just one
solution. This behavior is called {\em reentrant} \cite{simanek79}.

We conclude this Section observing that the phase
boundary line obtained within the mean-field approximation in the
path-integral approach is $1/\alpha=g(q,y)$ with
\cite{vanotterlo93,grignani00} \begin{equation}
\label{casodiag}
g(q,y)=\frac{\sum_n e^{-\frac{4}{y}(n+\frac{q}{2e})^2}
\frac{1}{1-4(n+\frac{q}{2e})^2} }
{\sum_{m}e^{-\frac{4}{y}(m+\frac{q}{2e})^2} }.
\end{equation}
Equation (\ref{casodiag}) coincides with Eqs. (\ref{r}) and (\ref{r0})
for $q=e$ and $q=0$, respectively.

\section{Capacitive Disorder}

In practical realizations of Josephson devices \cite{fazio01}, one has to
deal with disorder caused by offset charge defects in the
junctions or in the substrate \cite{krupenin00}.
Random offset charges cannot be made to vanish by using a gate for each
superconducting island since, in large arrays, too many electrodes would
be necessary, making impossible the cooling of the system at the desired
temperatures. In Ref. \cite{lafarge95} it was observed a sensible
variation ($\sim 40 \,\%$) of the resistance between the unfrustrated
and the fully frustrated array. Moreover, it may also
happen that the network's parameters are not uniform across the whole
array: despite recent advances in fabrication techniques, variation of
junction parameters associated to the shape of the islands can be also
of $20 \,\%$ \cite{fazio01}. Thus, it is relevant in many practical
situations to study JJA's also with randomly distributed self-capacitances:
this corresponds to have a random diagonal charging energy
\cite{mancini03,alsaidi03}.

In this Section, we shall determine the finite temperature 
phase diagram of JJA's with capacitive disorder 
(i.e., with random offset charges and/or random self-capacitances). 
To derive the phase boundary between the insulating and 
the superconducting phase, we shall use the mean-field approach 
for quantum JJA's with offset charges and diagonal capacitance 
matrices reviewed in the previous Section. One has to impose
now the self-consistency condition with a double average: the quantum
average and the one over the disorder.

As we shall see, charge disorder supports superconductivity;
furthermore, the relative changes of the insulating and
superconducting regions of the phase diagram depend crucially on the
weights of the $\delta$-like charge probability distribution. In the
physical relevant situation of two charge distributions peaked at the
values $q=0$ and $q=e$, increasing the frustrated weight favors the
superconducting phase. Also the randomness of the self-capacitances
leads to remarkable effects, namely, the superconducting phase increases
with respect to the case where disorder is not present. In the
following, we shall provide a quantitative analysis of these phenomena.

A pertinent extension of SCMFT in the presence of 
on-site disorder may be obtained introducing an order parameter 
averaged {\em also} over the disorder. In the following $
\langle \cdots \rangle$ denotes only the quantum average while 
$[\cdots]_{av}$ an average over the random variables. 
The single-site Hamiltonians of Eq. (\ref{MFA}) then become
\begin{equation}
H_{i}=-4E_{C}^{(i)}\frac{\partial ^{2}}{\partial
\varphi _{i}^{2}} -8iE_{C}^{(i)}\frac{q_{i}}{2e}\frac{\partial
}{\partial \varphi _{i}} -zE_{J} \left[\langle \cos {\varphi
}\rangle\right]_{av} \cos {\varphi _{i}}
\label{QPM_mf_random}
\end{equation}
where $E_{C}^{(i)}=e^2 C_{ii}^{-1}/2$. 
The Hamiltonian (\ref{QPM_mf_random}) depends on a random
variable $X$, which can be either $q_{i}$ or $E_C^{(i)}$. Thus, its
eigenfunctions and eigenvalues depend either on $q_{i}$ or $E_C^{(i)}$:
\begin{equation}
H_i \,\psi_n(\varphi_i;X)=E_n(X)\, \psi_n(\varphi_i;X) .
\end{equation}
The self-consistency condition is given by
 \begin{equation}
\left[\langle \cos {\varphi }\rangle\right]_{av}
 =\int d X \, P\left(X \right)
\frac{\sum_{n}e^{-\beta E_{n}(X)}\langle \psi _{n}(X)\vert\cos \varphi
\vert\psi _{n}(X)\rangle }{\sum_{n}e^{-\beta E_{n}(X)}}
 \label{self-con}
\end{equation}
where $P\left(X\right)$ is the probability distribution of $X$.

The phase boundary line is obtained from Eq. (\ref{self-con}) by
requiring $\left[\langle \cos {\varphi }\rangle\right]_{av}$ to be
small and by keeping only terms proportional to it. The self-consistency
condition yields a mean-field phase boundary line in agreement with the
results obtained by the path-integral approach \cite{mancini03}. The low
temperature behavior, obtained by a pertinent extrapolation of our
finite $T$ results, is consistent with the phase diagram obtained in
Ref. \cite{fisher89} (see Ref. \cite{mancini03}).

\subsection{Random Offset Charges}

In this Section we shall study JJA's at finite temperature with
random charge frustration. One assumes that the offset charges 
are independently distributed according to a probability distribution 
given by a sum of $\delta$-like distributions
\begin{equation}
P(q_{i})=\sum_n p_n \, \delta(q_{i}-ne)
\label{prob_delta_def}
\end{equation}
with $\sum_n p_n=1$. This corresponds to a random distribution of
charges which are integer multiples of $e$ and, actually, this is the
most realistic situation for a random distribution. Inserting the probability
distribution (\ref{prob_delta_def}) in Eq. (\ref{self-con}), one has
\be
\frac{1}{\alpha}= \int dq
\sum_n p_n \,\delta(q-ne) g(q,y)=\sum_{odd} p_n \,g(ne,y)+ \sum_{even} p_n
\,g(ne,y)
\label{boundary_line_charges}
\ee
where $\sum_{odd}$ ($\sum_{even}$) is a sum restricted to odd (even)
integer.
\begin{figure}
\includegraphics[height=.2\textheight]{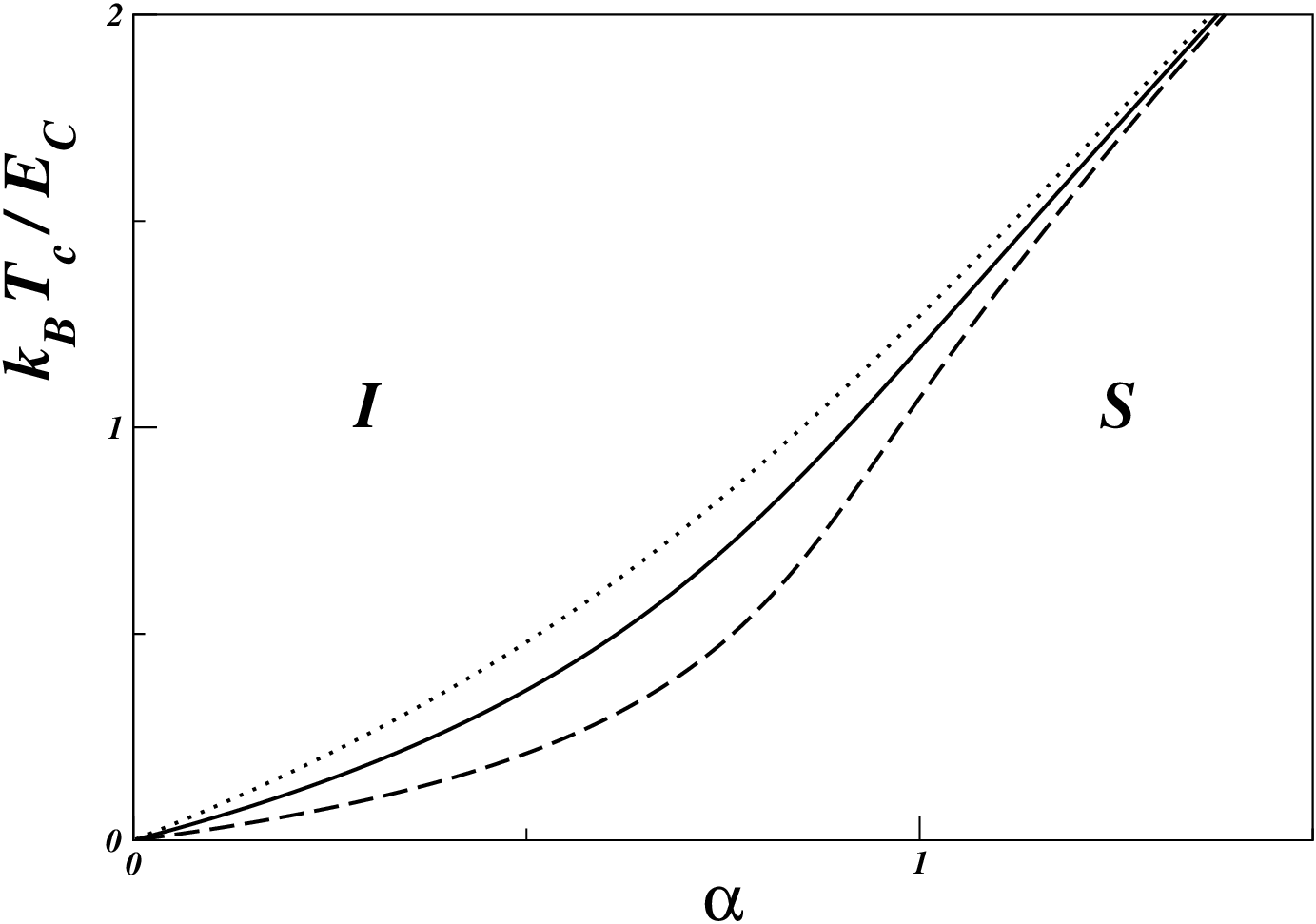}
\caption{Phase diagram for random offset charges with the
probability distribution given by Eq. (\protect\ref{prob_delta_def}).
In the plot we take $p_e=1/4$
(dashed line), $1/2$ (solid line), and  $3/4$ (dotted line).}
\label{fig2}
\end{figure}
Since the thermodynamical properties of the system are invariant under
the shift $q \to q+2ne$, one should note that $g(q+2ne,y)=g(q,y)$,
where $n$ is an integer. As a consequence, Eq.
(\ref{boundary_line_charges}) leads to
\begin{equation}
\frac{1}{\alpha}=p_0\, g(0,y)+p_e \,g(e,y)
 \label{prob_delta}
\end{equation}
where $p_0=\sum_{even} \, p_n$ ($p_e=\sum_{odd} p_n$) is the
probability that the offset charge $q$ is an even (odd) integer multiple
of $e$. The results obtained from Eq. (\ref{boundary_line_charges}) with
the probability distribution (\ref{prob_delta_def}) are displayed in
Fig. \ref{fig2}, where we plot the phase boundary line for $p_e=1/4,\:
1/2, \: 3/4$. One observes that increasing $p_e$ leads to an enlargement
of the superconducting phase. It is worth noting that applying the SCMFT
approximation with the probability distribution (\ref{prob_delta_def})
it is possible to find exactly the same result obtained in a
path-integral approach for a JJA model with diagonal capacitance
matrices \cite{mancini03}.

\subsection{Random Self-Capacitances}

\begin{figure}
\includegraphics[height=.2\textheight]{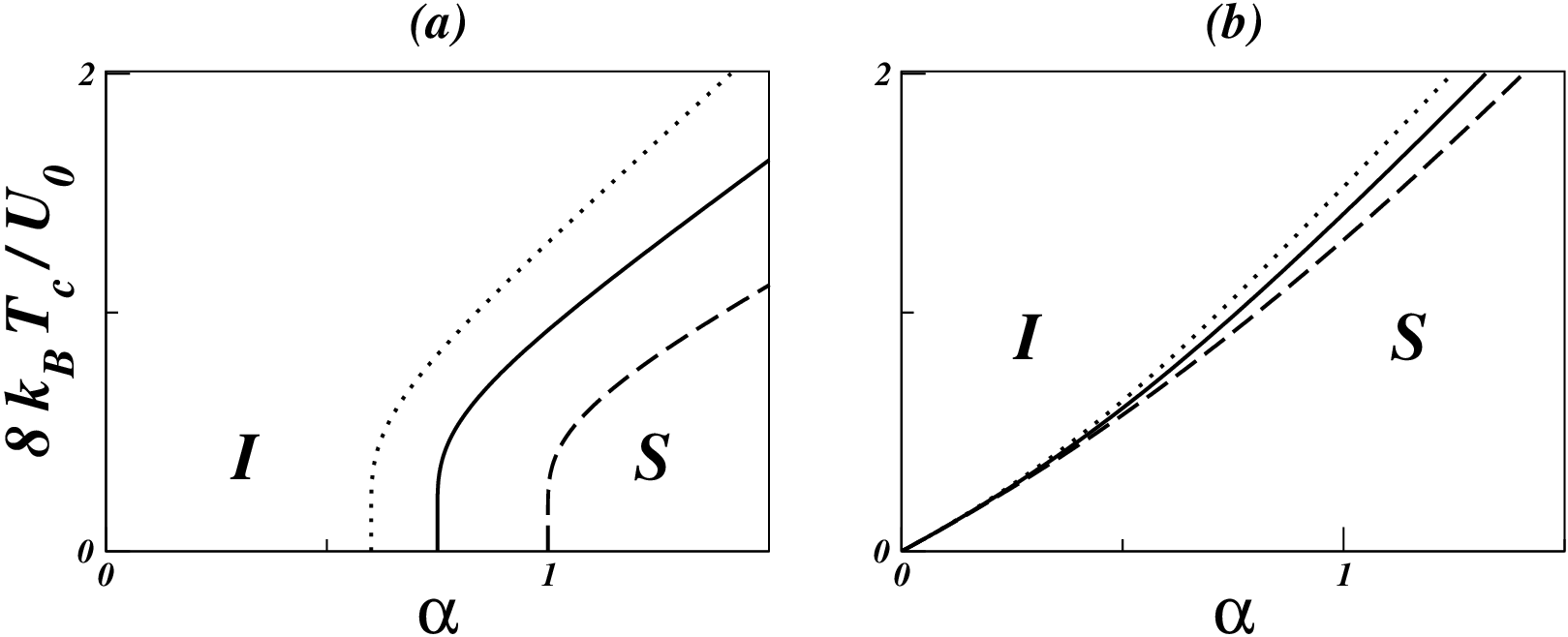}
\caption{Phase diagram in the $T_c-\alpha$ plane
for random diagonal charging energies distributed according to Eq.
(\protect\ref{prob_bimodal}) and uniform offset charge
$q=0$ (a), $q=e$ (b). $U_0$ is the average charging energy. We
take $U_1=1/2$ and $U_2=3/2$, while $p$ is $1/4$ (dashed line), $1/2$
(solid line) and $3/4$ (dotted line).
}
\label{fig3}
\end{figure}

In this Section we shall study the finite temperature phase diagram 
of JJA's with uniform charge frustration $q$ 
and random self-capacitances $C_{ii}$ 
with average $C_0$. Correspondingly, the average 
charging energy is $E_C^{0}=e^2/2C_0$.  It is useful to define the
charging energy terms $U_{ii}=8E_C^{(i)}$ and $U_0=8E_C^0$; averaging
the self-consistency equation (\ref{self-con}) over the random variables
$U_{ii}$, the equation for the phase boundary becomes
\begin{equation} \frac{1}{\alpha}=\int_{0}^{\infty} dU \:
\frac{P(U)}{U}\: g(q,U,y)
\label{bound_line_ec_random}
\end{equation}
where now  $\alpha=2 z E_J /U_0$, $y=8 k_B T_c / U_0$ and
$U=U_{ii}/U_0$. 
Equation (\ref{bound_line_ec_random}) can be also
obtained by using the path-integral approach \cite{mancini03}. The
function $g(q=e,U,y)$ is given by
\be
g(q=e,U,y)=
\frac{
\frac{4U+y}{4y} \: e^{-U/y}+\sum\limits_{n=1}^{\infty}
\frac{1}{1-4(n+1/2)^2} \: e^{-(4U/y)(n+1/2)^2}
}
{e^{-U/y}+ \sum\limits_{m=1}^{\infty} e^{-(4U/y)(m+1/2)^2}},
\ee
whereas the function $g(q=0,U,y)$ is given by
\be
g(q=0,U,y)=
\frac{
\sum\limits_{n=-\infty}^{\infty} e^{-4U/y}\frac{1}{1-4n^2}}
{\sum\limits_{m=-\infty}^{\infty} e^{-(4U/y)m^2}} .
\ee
If one, for instance, considers a bimodal distribution of the $U$'s,
then
\be
P(U)=p \: \delta(U_1-U)+(1-p) \: \delta(U_2-U),
\label{prob_bimodal}
\ee
where $U_1$ and $U_2$ are positive numbers. Inserting the
probability distribution (\ref{prob_bimodal}) in Eq.
(\ref{bound_line_ec_random}), one gets:
\begin{equation}
\frac{1}{\alpha}=p \, \frac{1}{U_1}\, g(q,U_1,y)+(1-p) \,\frac{1}{U_2}
\, g(q,U_2,y) .
\label{bound_line_bimodal}
\end{equation}
The phase boundary line given by Eq. (\ref{bound_line_bimodal}) is
plotted in Fig. \ref{fig3}. One observes that, when $q=0$, the
superconducting phase increases in comparison to the nonrandom case:
this is due to the factors $1/U_1$ and $1/U_2$ in Eq.
(\ref{bound_line_bimodal}), which make larger the contribution of
junctions with charging energies less than $U_{0}$. The increase of the
superconducting phase is thus due to a decrease of the effective value
of the charging energy. This phenomenon is largely independent from the
specific choice of the distribution \cite{mancini03}.

It is pertinent to observe that, when $q=e$ (maximum frustration induced
by the external offset charges), the randomness does not modify
considerably the phase diagram. This should be compared with the
unfrustrated case ($q=0$), where randomness sensibly affects the phase
diagram.

\section{Concluding Remarks}

In this paper, we reviewed the use of the self-consistent mean-field
theory to analyze the effects induced by offset charges on the finite
temperature phase diagram of Josephson junction arrays.

We reviewed, for a diagonal Coulomb interaction matrix, the explicit
derivation of the equation for the phase boundary line between the
insulating and superconducting phase. The resulting phase diagram is
drawn for a uniform offset charge distribution $q$: with $q=e$, the
superconducting phase increases with respect to $q=0$, and the model
exhibits superconductivity for all the values of $\alpha=z E_J/4E_C$. An
offset charge $q=e$ tends therefore to decrease the charging energy and
thus favors the superconducting behavior even for small Josephson energy
$E_J$.

Using a pertinent extension of the self-consistent mean-field
approach, we obtained here also the phase diagram at finite temperature
of JJA's with capacitive disorder. For a random distribution of offset
charges which are integer multiples of $e$ one has that the
superconducting phase increases in comparison with the unfrustrated
case. For arrays with random charging energies, the superconducting
phase increases with respect to the situation in which all
self-capacitances are equal.

It is comforting to see that our mean-field analysis 
provides results which are in very good agreement with 
those obtained by Quantum Monte Carlo simulations \cite{alsaidi03_2} 
for most of the phase diagram.

\section*{ACKNOWLEDGMENTS}
It is our pleasure to contribute with this paper to the volume to
honor the 60th birthday of Prof. Ferdinando Mancini. We are very glad
to have had the chance to benefit from many stimulating discussions
during the years of our friendship. We are grateful to F. Cooper,
G. Grignani, A. Mattoni, S. R. Shenoy and A. Tagliacozzo for
enlightening discussions. We acknowledge financial support by M.I.U.R.
through grant No. 2001028294.


%

\end{document}